\documentclass[journal,10pt]{IEEEtran}

\usepackage{cite,graphicx,amsmath,amssymb}
\usepackage{subfigure}
\usepackage{citesort}
\usepackage{fancyhdr}
\usepackage{mdwmath}
\usepackage{mdwtab}
\usepackage{balance}
\usepackage{xcolor}
\usepackage{bm}
\usepackage{amsthm}
\usepackage{threeparttable}
\usepackage{algorithm}
\usepackage{algorithmic}
\usepackage{multirow}
\usepackage{flafter}
\usepackage{makecell}
\usepackage{diagbox}
\usepackage{url}

\providecommand{\url}[1]{#1}
\allowdisplaybreaks
\begin{document}
\title{Reconfigurable Intelligent Surface-Aided Near-field Communications for 6G: Opportunities and Challenges}

\author{
Xidong~Mu,~\IEEEmembership{Member,~IEEE}, Jiaqi~Xu,~\IEEEmembership{Member,~IEEE}, Yuanwei~Liu,~\IEEEmembership{Fellow,~IEEE}, and Lajos~Hanzo,~\IEEEmembership{Fellow,~IEEE}

\thanks{Xidong Mu, Jiaqi Xu, and Yuanwei Liu are with the School of Electronic Engineering and Computer Science, Queen Mary University of London, London E1 4NS, UK, (email: \{xidong.mu, jiaqi.xu, yuanwei.liu\}@qmul.ac.uk).}
\thanks{Lajos Hanzo is with the School of Electronics and Computer Science, University of Southampton, Southampton SO17 1BJ, U.K. (e-mail: lh@ecs.soton.ac.uk).}
}

\maketitle
\begin{abstract}
Reconfigurable intelligent surface (RIS)-aided near-field communications is investigated. First, the necessity of investigating RIS-aided near-field communications and the advantages brought about by the unique spherical-wave-based near-field propagation are discussed. Then, the family of patch-array-based RISs and metasurface-based RISs are introduced along with their respective near-field channel models. A pair of fundamental performance limits of RIS-aided near-field communications, namely their power scaling law and effective degrees-of-freedom, are analyzed for both patch-array-based and metasurface-based RISs, which reveals the potential performance gains that can be achieved. Furthermore, the associated near-field beam training and beamforming design issues are studied, where a two-stage hierarchical beam training approach and a low-complexity element-wise beamforming design are proposed for RIS-aided near-field communications. Finally, a suite of open research problems is highlighted for motivating future research.
\end{abstract}

\section{Introduction}
With the rapidly spreading roll-out of the fifth-generation (5G) wireless network across the globe, researchers have turned their face to the sixth-generation (6G) concepts. One of the key objectives is to enhance the data rate, the latency, the space-air-ground coverage, and harness pervasive intelligence~\cite{6GWang}. To fulfill these stringent requirements, the transmission technologies have to be further enhanced, relying on the employment of extremely large multiple-input multiple-output (XL-MIMO) schemes, mmWave/terahertz (THz) bandwidths, satellite communications, just to name a few directions~\cite{6GWang}. Furthermore, hitherto unknown ground-breaking technologies should be conceived for revolutionizing the family of existing wireless technologies. Among these new technologies, reconfigurable intelligent surfaces (RISs) having a massive number of low-cost electromagnetic (EM) reflection elements constitute one of the most promising technologies for 6G. RISs can be deployed in wireless networks for beneficially ameliorating the signal propagation for much-needed signal enhancement and interference mitigation~\cite{Renzo2020JSAC}. More importantly, the near-passive full-duplex characteristics of RISs render them more attractive than active relaying technologies having expensive power-hungry radio frequency (RF) chains. Therefore, RIS-aided wireless communications pave the way for sustainable and ubiquitous 6G services. 

\begin{figure}[ht!]
    \centering
    \includegraphics[width= 3.5in]{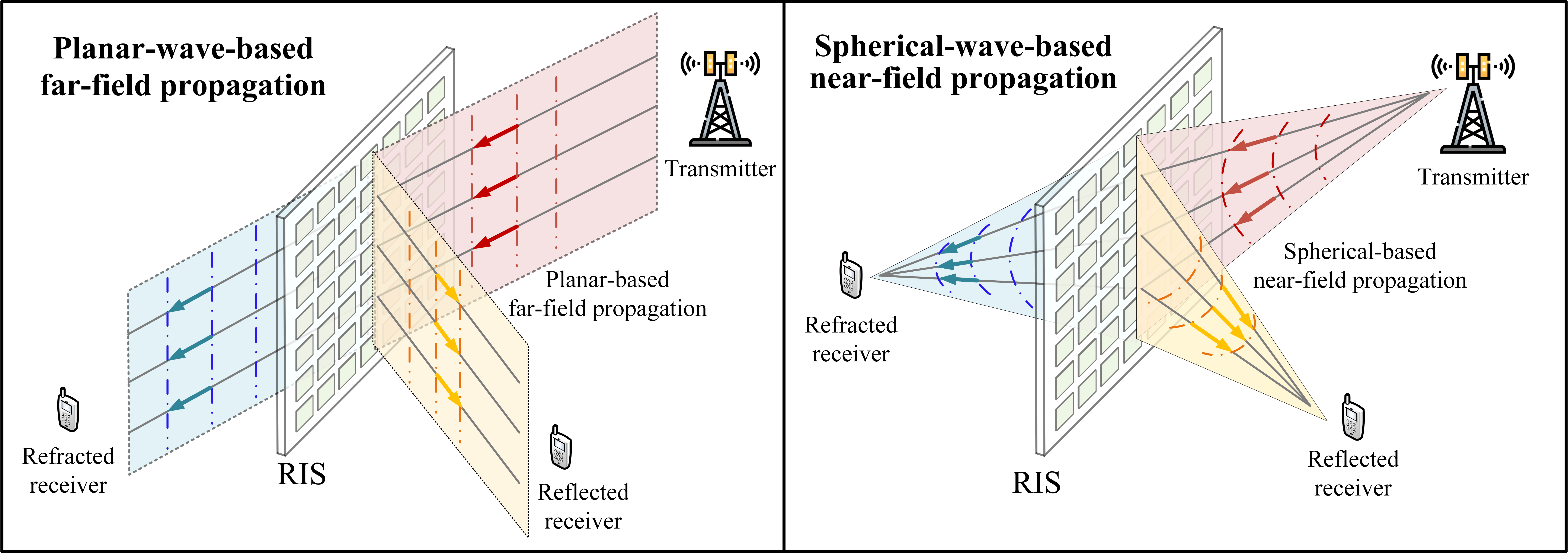}
    \caption{Illustration of RIS-aided far-field and near-field wireless communications, where the RIS can refract and/or reflect the incident signal to surrounding users.}\label{VS}
\end{figure}

Given the aforementioned benefits, extensive research efforts have been devoted to RIS-aided wireless communications, including but not limited to their passive beamforming coefficient design, channel estimation, and deployment location design~\cite{Renzo2020JSAC}. However, the majority of existing investigations relied on the assumption of far-field propagation, where both the incident and refracted/reflected wireless signals exhibited a planar wavefront, as shown in the left of Fig. \ref{VS}. By contrast, in RIS-aided near-field communications, the wavefront of wireless signals should be accurately characterized by their spherical wavefront, as shown in the right of Fig. \ref{VS}. 
\subsection{From Planar-wave-based Far-field Propagation to Spherical-wave-based Near-field Propagation}
Again, the signal propagation of RISs can be divided into near-field and far-field regions. One of the popular performance metrics of distinguishing the near-field and far-field regions is known as the Rayleigh distance, which is given by $\frac{2D^2}{\lambda}$~\cite{kraus2002antennas}. Here, $D$ and $\lambda$ denote the RIS aperture and the wavelength, respectively. According to~\cite{cui}, the cascaded transmitter-RIS-receiver channel falls within the near-field region if either the transmitter-RIS distance or the RIS-receiver distance is shorter than the Rayleigh distance. Although near-field signal propagation is not a new phenomenon, due to the low operating frequencies and small antenna apertures employed in the legacy wireless networks, the near-field coverage area was negligible. Since the far-field region was dominant, planar-wave-based far-field propagation has routinely been assumed for communication designs. However, when designing RISs for 6G, they will have a large aperture for realizing considerable array gains. They are eminently suitable for mmWave/THz communications for circumventing their blockage issues~\cite{Haiyang1}. Therefore, the near-field region in future RIS-aided communications becomes much larger and can no longer be ignored. For example, for an RIS with an aperture size of 1 meter employed at the frequency of 28 GHz, the resultant Rayleigh distance is approximately 187 meters.

\subsection{Advantages of Near-field Propagation for RISs}
Given the above discussion, considering RIS-aided near-field communications is not a choice but a necessity. Compared to conventional RIS-aided far-field communications, the unique spherical-wave-based near-field propagation leads to the following main advantages: 
\begin{itemize}  
  \item \textbf{High-rank Line-of-Sight (LoS) Channels}: Recall that RISs are generally harnessed for circumventing the lack of LoS channels between the transmitters and receivers. In RIS-aided far-field multi-antenna/multi-user communications, the LoS channels usually exhibit a low rank, they may even have a rank one. This limits the achievable spatial multiplexing gains of RISs. By contrast, the near-field propagation channels tend to exhibit high rank, even potentially full rank, which provides enhanced degrees-of-freedom (DoFs) in RIS-aided near-field communications~\cite{cui}.
  \item \textbf{Precise Near-field Beamfocusing Capability}: In contrast to far-field \emph{beamsteering}, where the energy of wireless signals is concentrated to a specific \emph{angle}, the spherical wavefront of RIS-aided near-field communications can be exploited to realize so-called \emph{beamfocusing}, where the energy of wireless signals is concentrated on a specific \emph{location} at a given \emph{angle and distance}. Given this new capability, the signal's energy in RIS-aided near-field communications can be more precisely delivered to the target receiver and prohibited from being leaked to other receivers, thus achieving efficient multi-user communications in the vicinity of RISs~\cite{Haiyang2}.
\end{itemize}

Since the investigation of RIS-aided near-field communications is still in its infancy, its benefits have not been widely recognized, Hence, this article aims for providing a systematic easy-reading introduction to RIS-aided near-field communications, including the near-field channel models, the fundamental near-field performance analysis, near-field beam training, and near-field beamforming design, complemented by a suite of future research challenges. The main contributions of this article can be summarized as follows.
\begin{itemize}
  \item We classify RISs into two categories, namely patch-array-based RISs and metasurface-based RISs. For each type of RISs, we discuss their key features, advantages, disadvantages, and near-field channel models. 
  \item We characterize the fundamental performance limits of RIS-aided near-field communications, namely their power scaling laws and effective DoFs (EDoFs). We highlight their performance differences wrt those in RIS-aided far-field communications.
  \item We discuss the associated beam training and beamforming design issues of RIS-aided near-field communications. In particular, a two-stage hierarchical beam training approach and an element-wise beamforming design are proposed, which significantly reduce the complexity of RIS-aided near-field communications.
\end{itemize}

\section{Category of RISs and Near-field Channel Models}
In this section, we will first introduce the family of patch-array-based and metasurface-based RISs. Then, we will discuss their respective near-field channel models.
\subsection{Patch-Array and Metasurface-based RISs}
Based on Fig. \ref{RIS}, we will introduce the key features, advantages, and disadvantages of these two types of RISs.
\subsubsection{Patch-Array-based RISs} Again, RISs are composed of numerous low-cost elements capable of reconfiguring their phase-shift and amplitude responses. In order to mitigate their grating lobes, these elements are typically spaced at half-wavelength intervals. We refer to these conventional RISs as patch-array-based RISs, as the phase-shift profile across the surface exhibits discontinuities. As illustrated in the left of Fig.~\ref{RIS}, each element in the patch-array-based RIS possesses its own unique phase-shift response. The phase-shift coefficient remains constant within an individual element and independently changes among the elements. Patch-array-based RISs are promising for employment at low frequencies, where the half-wavelength criterion suggests that the dimension of elements is on the order of a few centimetres~\cite{xu2022modeling}. Therefore, a number of low-cost positive-intrinsic-negative (PIN) diodes, varactor diodes, or delay lines can be installed at each element to realize high-resolution tunability. However, due to the short wavelength at high frequencies, the efficiency of patch-array-based RISs diminishes, hence their benefits might become marginal.

\subsubsection{Metasurface-based RISs} 
\begin{figure}[t!]
    \centering
    \includegraphics[width= 3.4in]{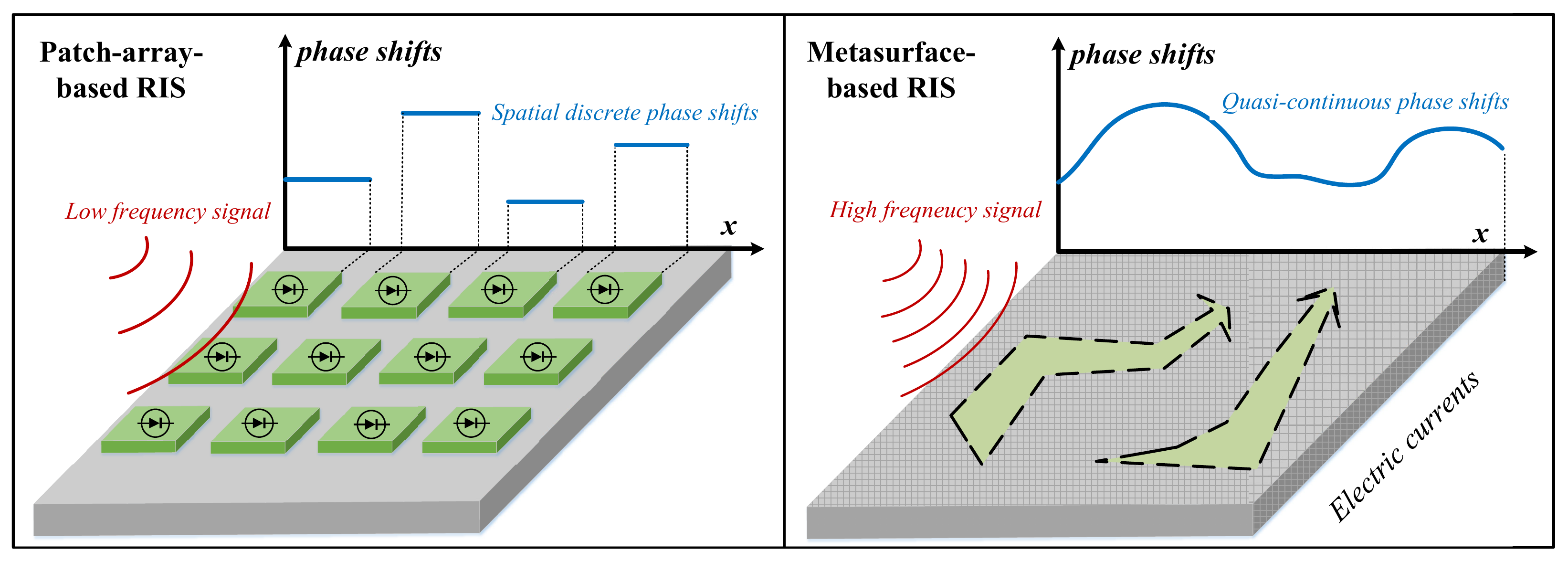}
    \caption{Illustration of patch-array-based RISs and metasurface-based RISs.}
    \label{RIS}
\end{figure}

Metasurface-based RISs are realized using massive periodic cells with dimensions ranging from a few millimetres to micrometres or even to the molecular scale~\cite{xu2022modeling}. As illustrated in the right of Fig.~\ref{RIS}, metasurface-based RISs have the capability of quasi-continuous operation across the entire surface. Consequently, metasurface-based RISs require advanced control of the EM properties, including their conductivity and permittivity. Compared to patch-array-based RISs, metasurface-based RISs are capable of operating at higher frequencies and exhibit beneficial phase-shift profiles, which leads to improved power scaling laws and EDoFs (see Section \ref{sec:per} for details). However, the sophisticated control of metasurface-based RISs increases both their optimization and implementation complexity.

\subsection{Near-field Channel Models}

Based on the above two types of RISs, we continue by introducing their representative near-field channel models for RIS-aided near-field communications.

\subsubsection{Patch-Array-based RISs}
For patch-array-based RISs, the RIS-aided near-field MIMO channel can be accurately represented using a channel matrix having dimensions equal to the number of transmit and receive antennas. The entry $\left( {i,j} \right)$ of the near-field channel is generally given by
\begin{align}\label{1}
{\left[ {{{\bf{G}}^{{\rm{LoS}}}}} \right]_{i,j}} = \sum_{m=1}^M{\beta _{i,j,m}}{e^{ - j\frac{{2\pi }}{\lambda } (\left\| {{{\bf{r}}_i} - {{\bf{s}}_m}}  \right\| + \left\| {{{\bf{s}}_m} - {{\bf{t}}_j}}  \right\| ) }},
\end{align}
where ${\beta _{i,j,m}}$ denotes the distance-dependent path loss of a cascaded link, $\bf{s}_m$ is the Cartesian coordinate of the $m$th RIS element, ${{\bf{r}}_i}$ and ${{\bf{t}}_j}$ represent the Cartesian coordinates of the $i$th receive antenna and the $j$th transmit antenna, respectively, while $\lambda$ denotes the wavelength. 
Observe that the near-field channel model in \eqref{1} has non-uniform channel gains and non-linear phases.

\begin{figure}[t!]
    \centering
    \includegraphics[width= 3.5in]{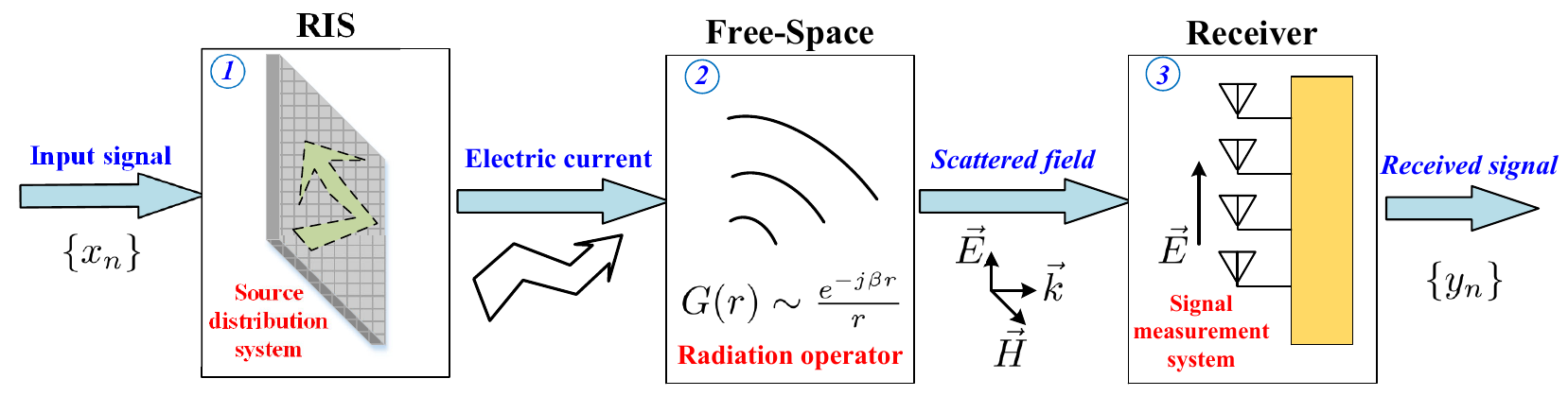}
    \caption{Three stages of the signal being refracted/reflected by the metasurface-based RIS to the receiver.}
    \label{Fig:op}
\end{figure}

\subsubsection{Metasurface-based RISs} For metasurface-based RISs, the channel modelling process is more complex due to their quasi-continuous phase profile. Note that the key distinguishing feature of metasurface-based RISs is the presence of surface electric currents. To demonstrate this, we characterize the incident signal refracted/reflected by metasurface-based RISs to the receiver in the following three stages with reference to Fig~\ref{Fig:op}.
\begin{itemize}
    \item \textit{From incident signals to electric currents}: In the first stage, the RIS can be regarded as a \textit{source distribution system} that produces a particular current density in space. When the wireless signals impinge upon the RIS, the incident EM field induces currents within the entire volume of RIS. 
    \item \textit{From electric currents to scattered fields}: In the second stage, the time-varying electric currents radiate a controlled scattered field. We refer to the mapping from electric currents to the scattered field as the \textit{radiation}. Note that the process of radiation is the key factor determining the achievable DoFs in the near field, which will be discussed in Section \ref{sec:dof}.
    \item \textit{From scattered fields to received signals}: In the third stage, the receiver can be regarded as a \textit{signal observation system}, where the strengths of the received electric field at different points on the aperture are calculated as a weighted sum. 
\end{itemize}

In the first stage of metasurface-based RISs, continuous electric current distributions should be adopted for capturing the EM responses instead of discrete phase-shift matrices. For the ideal scenario, where the receiver is completely accurate, the observation of the above third stage can be characterized by an identity matrix. In this case, the end-to-end channel gain between a transmitter and a receiver associated with an RIS source current $J$ is given by~\cite{xu2022modeling}:
\begin{equation}\label{A_222}
    |h(J)|^2 = \frac{GA_T}{4\pi d^2} \int_{V_T} J^*(\mathbf{s}_1) \int_{V_T} K(\mathbf{s}_1,\mathbf{s}_2, \mathbf{r})\mathrm{d}\mathbf{s}_1 \mathrm{d}\mathbf{s}_2,
\end{equation}
where $G$ is the directivity of the transmit antenna, $A_T$ is the aperture size of the RIS facing the transmitter, $d$ is the distance between transmitter and RIS, $\mathbf{s}_1$, $\mathbf{s}_2$ are source points within the metasurface-based RIS, $V_T$ denotes the volume of the metasurface-based RIS, $K(\mathbf{s}_1,\mathbf{s}_2, \mathbf{r})$
is the kernel function of the radiation operator, and $\mathbf{r}$ denotes the coordinates of the receiver. Note that the end-to-end near-field channel gain in \eqref{A_222} is subject to a given normalized source current distribution $J^*(\mathbf{s})$. In practice, the current distribution has to be optimized harnessing specific near-field receivers.

\section{Fundamental Performance Limits of RIS-aided Near-field Communications}\label{sec:per}

Having highlighted the near-field channel models of RISs, in this section we characterize the fundamental performance limits of RIS-aided near-field communications and discuss the performance differences both between the near-field and far-field regions as well as between both types of RISs.

\subsection{Power Scaling Law}
\begin{figure}[t!]
    \centering
    \includegraphics[width= 2.8in]{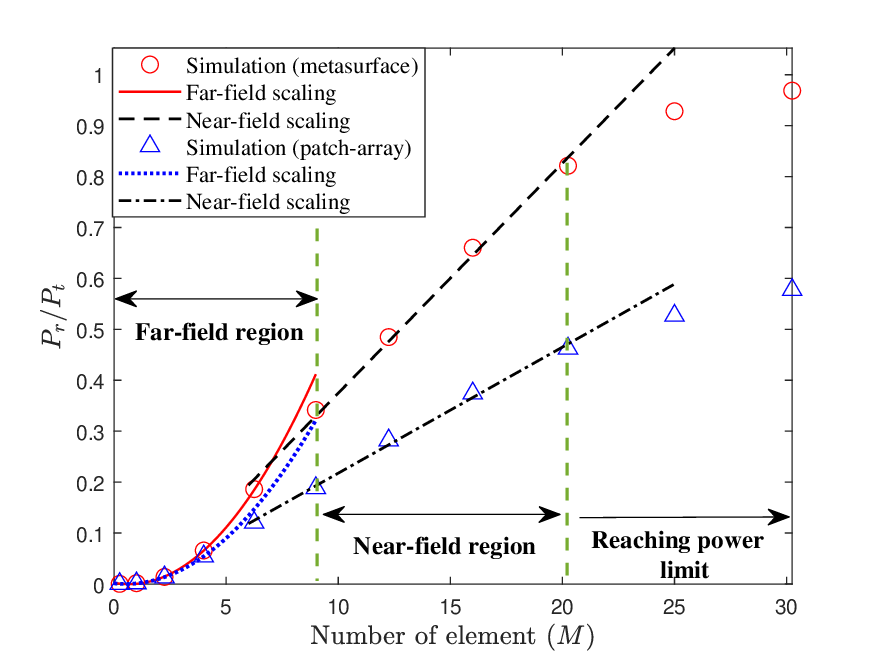}
    \caption{Power scaling laws for patch-array-based and metasurface-based RISs. The receiver is located at a fixed distance of $d=10$ meters with respect to the centre of RIS and the dimension of the receiver is $z_R \approx 5.0$ cm. The number of patch-array-based RIS elements ranges from $1$ to $20$ and the aperture size of the metasurface-based RIS ranges from $1$ cm$^2$ to $4$ cm$^2$. The wavelength of the carrier signal is $0.01$ m.}
    \label{Fig:power}
\end{figure}

In the context of our discussion, the power scaling law is defined as the ratio between the received power $P_r$ and transmit power $P_t$, which is a function of the RIS size.
To study the power scaling differences between RIS-aided near-field communications and far-field communications, we consider a single-user system, where the RIS phase-shift coefficients are configured according to the co-phasing condition for maximizing the end-to-end channel gain. We first discuss the different power scaling laws of RIS-aided near-field communications compared with those of RIS-aided far-field communications. Then, we discuss the power scaling law differences between patch-array-based and metasurface-based RISs in near-field communications. In the following, we assume plane wave excitation. When the positions of the transmitter, RIS, and receiver are fixed, the received power is upper-bounded by the transmit power, regardless of how large the RIS may become~\cite{EB}. Thus, we consider near-field power scaling before this limit is reached.

\subsubsection{RIS-aided far-field communications vs. near-field communications}
Observe in Fig. \ref{Fig:power} that conventional RIS-aided far-field communications exhibit \emph{quadratic} power scaling. In particular, for patch-array-based RISs, the received power scaling laws are proportional to the square of the number of RIS elements~\cite{Renzo2020JSAC}. By contrast, for metasurface-based RISs, the received power scales with the square of the RIS aperture size~\cite{xu2022modeling}. However, for RIS-aided near-field communications, the received power scales \emph{linearly} with the number of RIS elements or with the RIS aperture size. This is because within the near-field region, the subchannels associated with each of the RIS elements have different gains due to the varying path lengths.
Therefore, the ratio of received power and transit power ($P_r/P_t$) of RIS-aided near-field communications scales slower than in RIS-aided far-field communications. To further illustrate this phenomenon, Fig.~\ref{Fig:power} portrays $P_r/P_t$ for both patch-array-based and metasurface-based RISs, where the receiver has a fixed distance from the RIS. Observe in Fig. \ref{Fig:power} that as we increase the number of patch-array-based RIS elements and the metasurface-based RIS aperture size, the near-field propagation becomes dominant and the corresponding $P_r/P_t$ scaling gradually changes from the rapid parabolic-shaped far-field quadratic scaling to the moderate linear near-field scaling.

\subsubsection{Patch-Array-based RISs v.s. Metasurface-based RISs}
For patch-array-based and metasurface-based RIS-aided near-field communications, there is a slight difference in power scaling $P_r/P_t$. 
This is because for patch-array-based RISs, each element is only able to perform a uniform phase-shift value over its cross-section. By contrast, metasurface-based RISs facilitate more advanced control of the quasi-continuous phase-shift coefficients across the aperture.
As a result, a higher normalized received power $P_r/P_t$ can be achieved for metasurface-based RISs than for patch-array-based RISs~\cite{xu2022modeling}. This is also confirmed by Fig.~\ref{Fig:power}, where the $P_r/P_t$ slope of patch-array-based RISs is less steep than that of metasurface-based RISs.

\subsection{Effective Degrees-of-Freedom}\label{sec:dof}
The EDoFs of a channel represent the number of independent parallel sub-channels that can be supported.
The achievable EDoFs are determined by the hardware capabilities of the RIS and transceivers as well as by the positioning of the receiver. 
In the following, we discuss and compare the EDoFs of the RIS-to-receiver LoS MIMO channel in near-field communications and far-field communications, noting that the EDoFs of the LoS multi-antenna-transmitter-to-RIS channel can be characterized similarly. Generally speaking, the EDoFs of the end-to-end RIS-aided channel are upper-bounded by the minimum of the EDoFs of the transmitter-to-RIS and RIS-to-receiver channels.
\subsubsection{RIS-aided Far-field Communications vs. Near-field Communications}
For RIS-aided far-field communications, the EDoF of the RIS-to-receiver LoS MIMO channel is strictly equal to one. By contrast, for RIS-aided near-field communications, the RIS-to-receiver LoS MIMO channel exhibits higher EDoFs, which are generally larger than 1. This is because the varying path lengths associated with non-linear phases allow the near-field channel to have a higher rank. 
However, the near-field EDoFs further depend on the shape, size, and geometric orientations of the transceivers. Nevertheless, \cite{volume} shows that the maximum numbers of EDoFs between two rectangular prism volumes are roughly
\begin{equation}
    N_{\text{max}} = \frac{V_RV_T}{4(\lambda r)^2\Delta z_T \Delta z_R},
\end{equation}
where $V_{R/T}$ are the volumes of the receiver/RIS, $r$ is the distance, $\Delta z_{R/T}$ are the width of the receiver/RIS. As we can see in this formula, the EDoFs increase with the volumes and with the decrease of distance $r$. As a result, as the receiver moves into the near-field regime of an RIS, the available EDoFs may soon exceed the number of elements a patch-array RIS may have. However, for the metasurface-based RIS, it can fully exploit this $N_{\text{max}}$. In the following, we provide a detailed discussion of the near-field EDoFs for patch-array-based and metasurface-based RISs.
\subsubsection{Patch-Array-Based RISs vs. Metasurface-Based RISs}
For patch-array-based RIS, the EDoF of the RIS-user channel is equal to its effective rank~\cite{edof}. 
Consequently, the maximum achievable EDoFs are limited by the number of RIS elements and the number of transceiver antennas. This implies that even when the near-field radiation operator exhibits higher EDoFs, this cannot be achieved due to the hardware limitation of patch-array-based RISs. However, for metasurface-based RISs, the near-field channel given in \eqref{A_222} is not restricted to a finite-dimensional matrix form. As a benefit, metasurface-based RIS-aided near-field communications generally have higher EDoFs than the same-sized patch-array-based RIS. To analytically derive the EDoFs of the near-field channel for metasurface-based RISs, the EDoFs of the scattered field (as illustrated in Fig.~\ref{Fig:op}) have to be exploited \cite{added_dof_reference}. For the scenario, where the transceiver is equipped with a large number of antennas or a continuous-aperture antenna, there is no restriction on the EDoFs of metasurface-based RISs and the maximum EDoFs determined by the near-field physical channel may indeed be achieved. Existing study~\cite{1589439} shows that \textit{the EDoFs of a near-field channel is given by the number of large singular values of the radiation operator}. For metasurface-based RISs, this number depends on the geometries of the RIS and the receiver. For the case where the apertures of RIS and receiver are parallel to each other, the EDoFs between a metasurface-based RIS and a receiver are proportional to $S/r^2$, where $S$ is the aperture size of the metasurface-based RIS and $r$ is the communication distance~\cite{xu2022modeling}. 

The power scaling laws and EDoFs of RIS-aided far-field communications and near-field communications are summarized and compared in Table~\ref{tab:d}. In general, metasurface-based RIS-aided near-field communications achieve a better performance than patch-array-based RIS-aided near-field communications.

\begin{table*}[h!]
\caption{Performance differences between RIS-aided far-field communications and near-field communications.}
\centering
\label{tab:d}
\resizebox{\textwidth}{!}{
\begin{tabular}{|l|r|r|r|r|}
\hline
Performance metrics       & RIS-aided far-field communications & RIS-aided near-field communications & Patch-array-based RIS-aided near-field communications & Metasurface-based RIS-aided near-field communications                               \\ \hline
Power scaling law    & Quadratic scaling     & Linear scaling  &  Small scaling slope & High scaling slope      \\ \hline
Transmitter/Receiver-RIS LoS MIMO channel EDoFs & Strictly equal to 1 & Usually larger than 1  & Limited by element/antenna number  & Not limited\\ \hline
\end{tabular}
}
\end{table*}

\section{Beam Training and Beamforming Design for RIS-aided Near-field Communications}
Having discussed the fundamental performance limits of RIS-aided near-field communications, in this section, we discuss the specifics of efficient beam training and beamforming design of RIS-aided near-field communications.  

\subsection{Two-stage Hierarchical Near-field Beam Training}
\subsubsection{Overview of Near-field Beam Training} The accurate acquisition of channel state information (CSI) is of vital importance for communication design. However, for RIS-aided near-field communications, CSI estimation becomes quite challenging. This is because having a large number of RIS elements and transceiver antennas leads to high-dimensional channels. The corresponding complexity of estimating the complete CSI may become excessive. Moreover, the passive nature of RISs makes the challenge more grave. As a remedy, beam training is proposed. Relying on a predefined codebook comprised of different passive beamforming elements representing specific angular beams, beam training aims for determining the specific location of the target receiver both in terms of its angle and distance by selecting the optimal passive beamformer that achieves the maximum received signal power. By doing so, a high-quality initial link can be established for reduced-dimensional CSI estimation of the cascaded BS-RIS-user channel.

Although there are numerous research contributions on far-field beam training, they are unsuitable for near-field beam training. This may be explained from both a \emph{codebook design} and \emph{training protocol} perspective, as follows. The codebook design is the core of beam training, which directly determines the ultimate accuracy. Upon recalling that the near-field channels require both angular and distance information, the angle-only codebooks developed for far-field beam training become inefficient due to the \emph{energy-spread} effect \cite[Section IV-A]{chenyu} Explicitly, the energy of a far-field beamformer designed for a specific angle will spread to a range of different angles in the near field. As a result, the accuracy of the beamformer becomes gravely degraded. Circumventing this requires a codebook tailored for near-field beam training, where the codebook has to cater for both the discrete angular domain and the discrete distance domain, i.e., in the polar domain. Hence, compared to the far-field codebook, the near-field codebook becomes much larger, which further increases the complexity of near-field beam training. This calls for efficient near-field training protocols to be developed. 
\begin{figure}[t!]
    \centering
    \includegraphics[width= 2.6in]{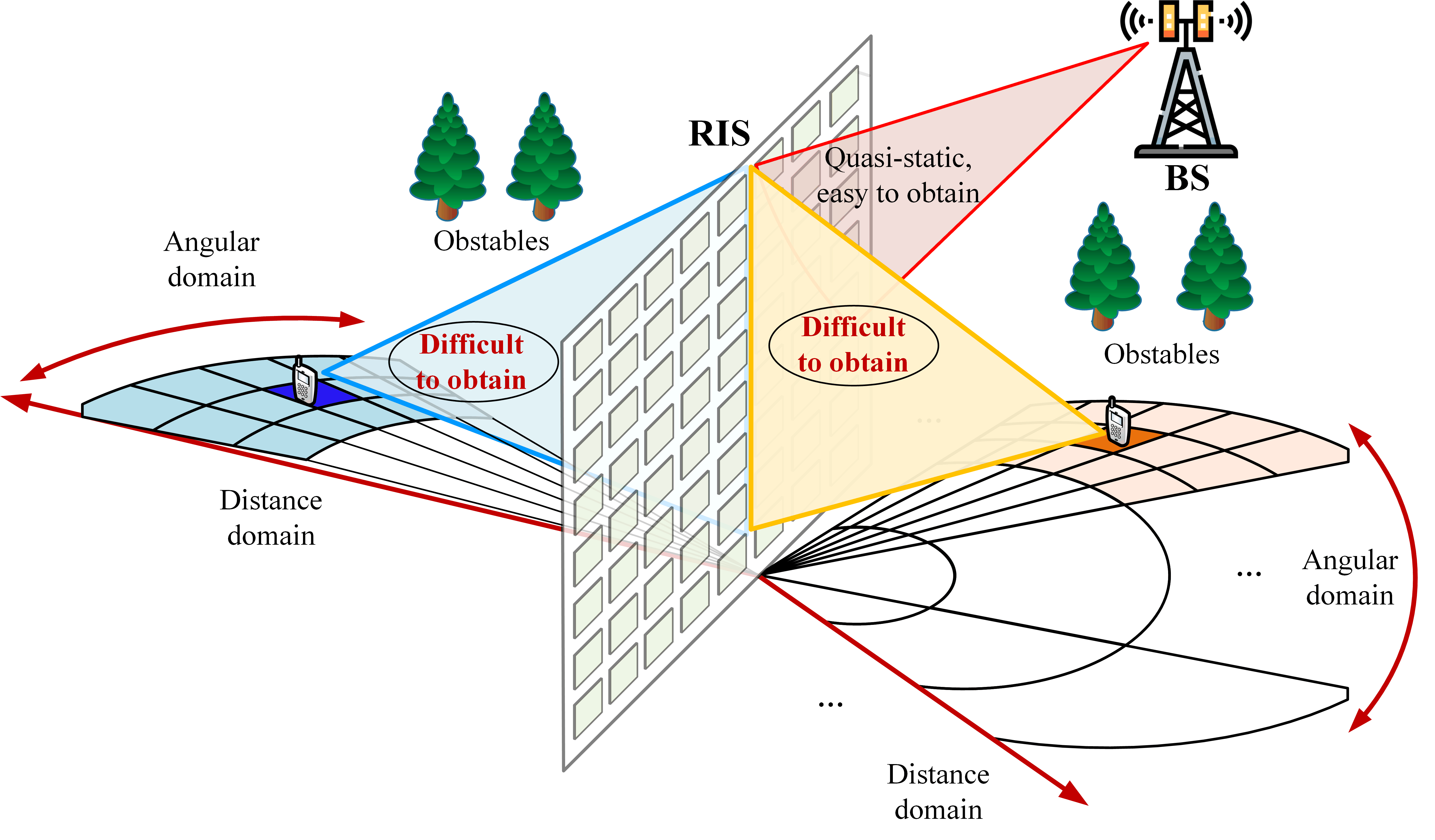}
    \caption{Illustration of the beam training in RIS-aided near-field communications.}
    \label{beamtraining}
\end{figure}

\subsubsection{Proposed Beam Training Approach} Given the above challenges, a novel two-stage hierarchical beam training approach was proposed in \cite{chenyu}, which can be harnessed for RIS-aided near-field communications. As shown in Fig. \ref{beamtraining}, the BS and RIS usually have fixed locations and thus have a LoS link between them. The corresponding BS-RIS channel is typically quasi-static in practice. The main challenge lies in the beam training of the RIS-user near-field transmitting/reflecting channel. To address this issue, based on the BS-RIS channel obtained, the BS's transmit beamforming can be directed towards the RIS. Then, the cascaded BS-RIS-user near-field beam training can be reduced to the RIS-user near-field beam training relying on the proposed two-stage hierarchical beam training approach. In the following, we provide a brief introduction to the proposed beam training approach.

\begin{figure}[t!]
    \centering
    \includegraphics[width= 2.8in]{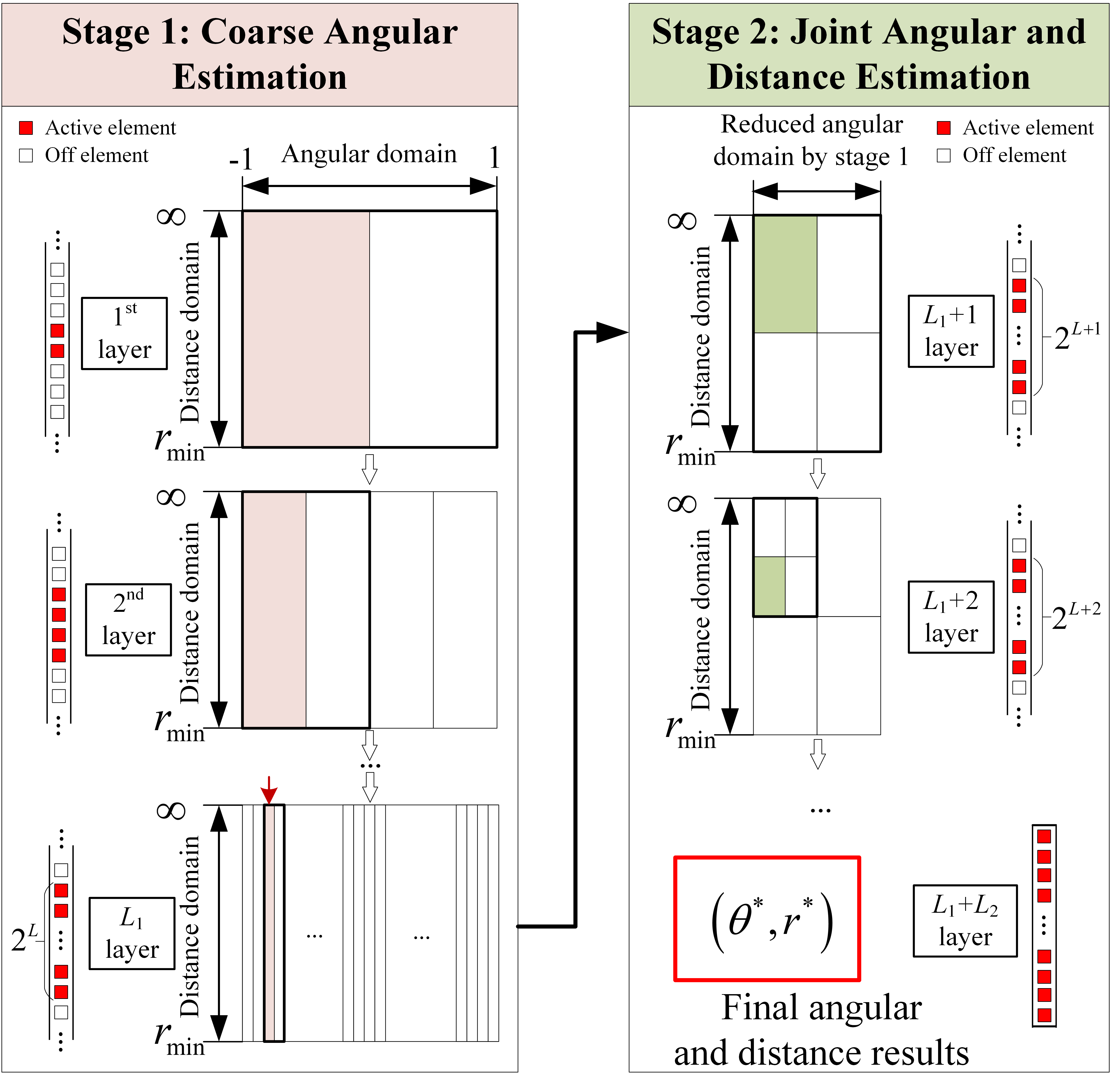}
    \caption{The proposed two-stage hierarchical beam training approach with the sub-array technique. More technical details can be found in \cite{chenyu}.}
    \label{2stage}
\end{figure}

To mitigate the above-mentioned near-field energy-spread effect, the proposed approach relies on initially activating a small central sub-array of the entire array during the beam training. The key idea is that upon activating a small sub-array, far-field propagation becomes dominant and the near-field energy-spread effect is reduced. Motivated by this observation, as shown in Fig. \ref{2stage}, two stages are involved in the proposed approach, namely the coarse angular estimation stage of far-field communications, followed by the joint angular and distance estimation stage. The number of layers in stage 1 and stage 2 are denoted by $L_1$ and $L_2$, respectively. With the gradual increase of the number of layers, the number of activated RIS elements also increases until all elements become activated in the last layer. The total number of layers, $L_t=L_1+L_2$, is determined by the total number of RIS elements denoted by $N$, i.e., $L_t={{{\log }_2}\left( N \right)}$. For each stage, the main procedure may be formulated as follows: 
\begin{itemize}
    \item \textbf{Stage-1 Coarse Angular Estimation}: As shown at the left of Fig. \ref{2stage}, each layer only activates a relatively small number of RIS elements. In this case, the users can be regarded to be located in the far-field region of the RIS, where the near-field energy-spread effect is negligible. Therefore, the conventional hierarchical far-field codebook that only distinguishes the angular domain is employed for coarsely estimating the user's angular information. 
    \item \textbf{Stage-2 Joint Angular and Distance Estimation}: As shown at the right of Fig. \ref{2stage}, given a reduced user angle candidate set estimated by stage 1, this stage further increases the number of activated RIS elements, hence near-field propagation becomes dominant. This stage then employs an appropriately designed hierarchical polar-domain codebook for jointly estimating the user's angular and distance information. In particular, each codeword within the hierarchical polar-domain codebook corresponds to a distinct combination of a sampled angular direction and a sampled distance. The design of hierarchical polar-domain codebooks follows two criteria. (1) In each layer, the entire angular and distance domains (i.e., the polar domain) are comprehensively represented by all the codewords. (2) The polar domain of a codeword within a particular layer should encompass the union of polar domains represented by several codewords in the subsequent layer.
\end{itemize}
The main advantage of the proposed two-stage hierarchical beam training technique is that the training overhead can be significantly reduced. It was shown in \cite{chenyu} that the training overhead in the proposed approach (in the order of ${\cal O}\left( {{{\log }_2}\left( N \right)} \right)$) reduces to 1\% and 5\% of those in the exhaustive search-based near-field beam training (in the order of ${\cal O}\left( {NS} \right)$) and in the two-phase beam training of \cite{Zhang} (in the order of ${\cal O}\left( {N+S} \right)$), respectively, where $S$ denotes the number of sampled distances. Note that there is a trade-off between the training overhead and the achievable performance in the proposed two-stage hierarchical beam training approach for different $L_1$ and $L_2$. Due to the employment of the polar-domain codebook, each layer in stage 2 can achieve a higher accuracy but consumes a higher training overhead than that in stage 1 using the angular-domain codebook. Therefore, the $L_1$ and $L_2$ should be carefully selected to strike a good trade-off selection. Detailed discussions of selecting $L_1$ and $L_2$ can be found in ~\cite[Section IV-B]{chenyu}.

\subsection{Low-complexity Element-wise Near-field Beamforming Design}
Efficient beamforming design is critical for creating a ``smart radio environment'' by RISs and thus for improving the overall communication performance. For RIS-aided near-field communications, an intrinsic feature is that the total number of RIS elements is generally large, which leads to potentially high computational complexity. Therefore, the existing beamforming methods proposed for RIS-aided far-field communications might be unsuitable for deriving high-dimensional near-field beamforming solutions. To address this issue, a novel element-wise optimization framework was proposed for simultaneously transmitting and reflecting (STAR)-RIS-aided near-field communications in \cite{haochen}. For a STAR-RIS having $N$ elements, the key idea is that only a single specific STAR-RIS coefficient is optimized at a time, with all the other $N-1$ coefficients fixed. Each element-wise STAR-RIS coefficient optimization problem is convex and easy to solve. Therefore, the resultant computational complexity increases only \emph{linearly} with the number of STAR-RIS elements, i.e., in the order of ${\cal O}\left( {N} \right)$, compared to that of the conventional beamforming method in the order of ${\cal O}\left( {N}^3 \right)$. This is attractive for practical implementations. In \cite{haochen}, it was shown that the computational time of the proposed element-wise approach required for optimizing the entire STAR-RIS beamforming vector in each iteration is merely 0.04\% of that of the conventional optimization approach. The performance gap in terms of the weighted sum rate between the proposed approach and the conventional optimization approach is less than 5\%.

\begin{figure}[ht!]
    \centering
    \includegraphics[width= 2.8in]{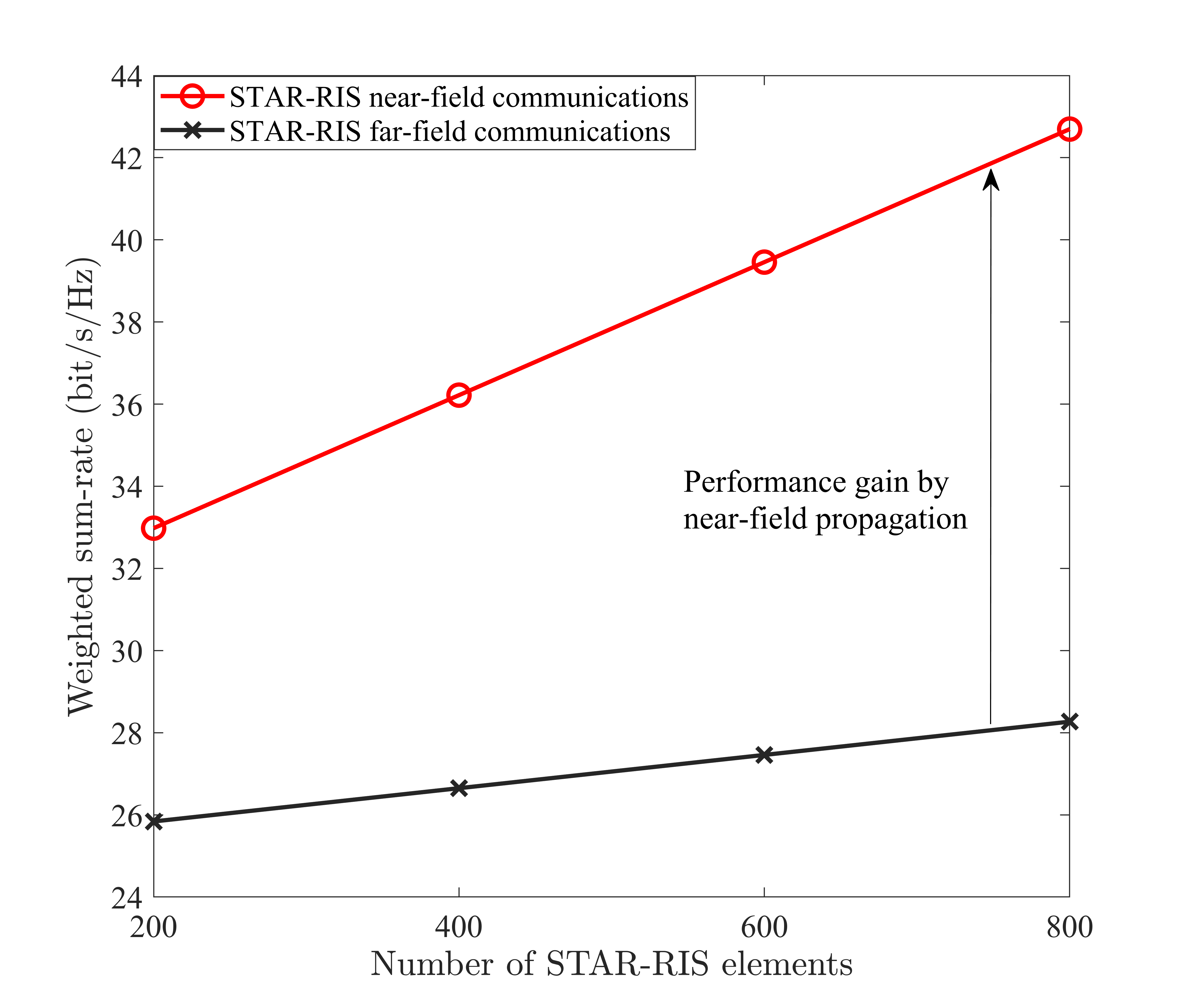}
    \caption{Achieved weighted sum rate versus the number of STAR-RIS elements in near-field communications and far-field communications. The system parameters can be found in \cite{haochen}.}
    \label{BF}
\end{figure}

To further illustrate the benefits of RIS-aided near-field communications, Fig. \ref{BF} compares the performance of STAR-RIS-aided multi-user MIMO communication in the near-field and far-field propagations \cite{haochen}. Observed that near-field propagation attains a higher weighted sum rate than far-field propagation. This is because the high-rank LoS channel and the near-field beamfocusing capability further mitigate the multi-user interference and thus improve the data rate. Moreover, it can be found that the rate enhancement attained by increasing the number of STAR-RIS elements is more pronounced in the near field than in the far field. This also underscores the benefits of RIS-aided near-field communications.

\section{Conclusions and Research Opportunities}
In this article, RIS-aided near-field communications has been discussed for future 6G wireless networks. In particular, both patch-array-based and metasurface-based RISs were discussed and the corresponding near-field channel models were introduced. Based on the near-field channel models, the power scaling law and EDoFs were analyzed, which are compared to those of RIS-aided far-field communications. Furthermore, the beam training and beamforming design of RIS-aided near-field communications were discussed. A two-stage hierarchical near-field beam training approach was proposed, which is capable of significantly reducing the training overhead. An element-wise beamforming design method was developed for obtaining the high-dimensional beamforming coefficients at a low computational complexity. Note that the design of RIS-aided near-field communications still has many research challenges, some of which are discussed as follows:

\begin{itemize}
    \item \textbf{Near-field CSI estimation, beam training, and beamforming design for metasurface-based RISs}: While metasurface-based RISs hold significant promise for performance enhancement in near-field communications, the intricate nature of the Green's function-based near-field channel model and the quasi-continuous configuration introduce more design challenges compared to patch-array-based RISs. Further research efforts, involving advanced theories and mathematical tools, are required to fill the knowledge gap.
    \item \textbf{Dynamic RIS configuration for adjustable near-field and far-field regions}: Since the Rayleigh distance depends on the RIS aperture size, adjustable near-field and far-field regions can be facilitated by switching specific RIS elements on and off, i.e., modifying the aperture size. Note that although higher DoFs and capacity enhancements can be achieved in the near-field region, the complexity of determining the CSI and beamforming design significantly increases. By exploiting the adjustable near-field and far-field regions provided by the dynamic RIS configuration, an appealing tradeoff can be struck between the performance achieved and the complexity imposed. This constitutes an interesting future research topic.
    \item \textbf{Exploiting generative artificial intelligence (GAI) in RIS-aided near-field communications}: GAI techniques have demonstrated substantial potential in the field of wireless communications. To address the complex optimization challenges associated with near-field channels and the vast number of optimization variables in RIS-aided near-field communications, GAI techniques, such as generative adversarial networks, transformers, and diffusion models, are emerging as promising solutions. For instance, their generative capabilities can be harnessed to create suitable codebooks and efficient beam search strategies during near-field beam training, as well as to facilitate near-field beamforming design and resource management in the presence of CSI uncertainties. These applications require further investigation.
\end{itemize}

\bibliographystyle{IEEEtran}

\end{document}